\documentclass[12pt,a4paper]{article}
\usepackage[T1]{fontenc}
\usepackage{dfttrob}
\usepackage[standard]{refstyle}
\usepackage{latexuseful2e}
\usepackage{amsmath}
\usepackage{epsfig}
\dfttnum{FISIST/19-98/CENTRA\\December 15, 1998}

\def\Smax{S_{\text{max}}}
\def\Pp{P_{\text{perc}}}
\def\fp{f_{\text{perc}}}

\begin{document}

\title{Percolation approach to phase transitions \\ in
high energy nuclear collisions}
\author{A. Rodrigues, R. Ugoccioni and J. Dias de Deus\\[3pt]
 \small\it CENTRA and Departamento de Física (I.S.T.),\\ 
 \small\it Av. Rovisco Pais, 1096 Lisboa Codex, Portugal}
\maketitle

\begin{abstract}
We study continuum percolation in nuclear collisions for
the realistic case in which the nuclear matter distribution
is not uniform over the collision volume,
and show that the percolation threshold is 
increased compared to the standard, uniform situation.
In terms of quark-gluon plasma formation this 
means that the phase transition threshold is pushed to higher 
energies.
\end{abstract}
\newpage

\section*{}
One of the main aims of the study of relativistic nucleus-nucleus
collisions is to find signs of deconfined partonic matter, known as
quark-gluon plasma (QGP), which is expected to form at large
energy densities \cite{QGP:review}.  
Many calculations have been proposed in order to estimate the
threshold for QGP formation; in particular the
idea of using percolation techniques has been explored in the past
\cite{perc:previous} and has been more recently applied
to the high energy data available today, namely the NA50
results on Pb--Pb collisions at the SPS collider \cite{NA50:recent}, 
which reported observation of strong, anomalous suppression of the $J/\psi$
signal. 
The authors of \cite{Pajares:perc,NardiSatz} used the percolation
approach to support the explanation of NA50 results in terms of QGP 
formation.
In this paper we will apply this technique to the realistic case in
which the nuclear matter is not uniformly distributed in the collision
volume.

\begin{figure}
  \begin{center}
  \mbox{\epsfig{file=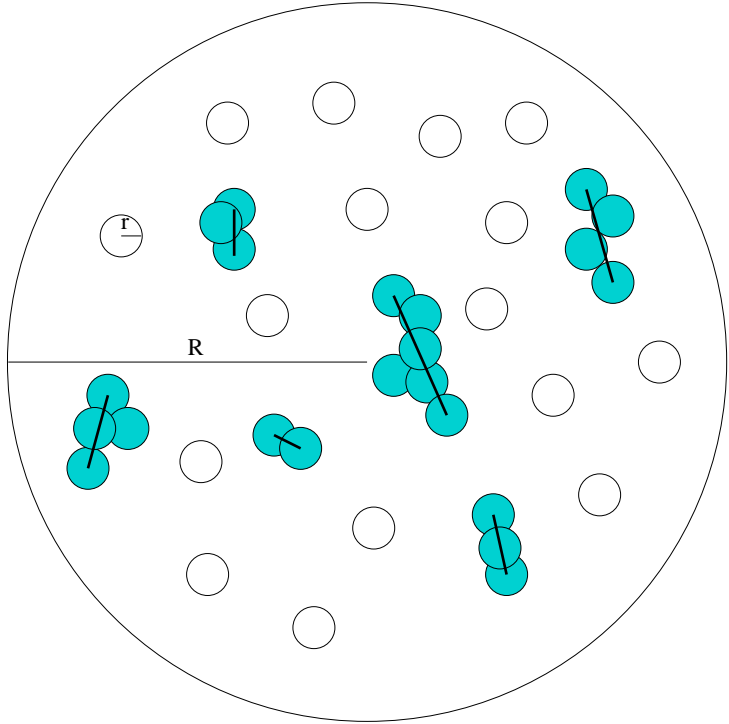,width=0.6\textwidth}}
  \end{center}
  \caption[discs]{Sketch of a cross section of a nuclear collision.
  Discs represent ``strings'' of colour exchange; clusters of fused
  strings are indicated by shading; within each cluster, the thick line
  indicates its diameter.}\label{fig:discs}
  \end{figure}

In the collision of two nuclei of relativistic energy, each
nucleon is subject to several interactions; these interactions are
often described %\cite{AA:models},
after the additive parton model, by colour exchange
between pairs of quarks (colour triplets) so that the collision
results in a large number of ``colour strings'' stretched between
pairs of partons, the pairs being distributed over the interaction
volume.
If we consider central collisions between identical nuclei, a cross
section in the transverse plane will show in the above framework
small discs (the cross
section of the strings, indicating colour exchange taking place) 
scattered randomly over a larger circular area,
representing the size of the colliding nuclei at zero impact parameter
(see Figure~\ref{fig:discs}). 
The number of strings in these models grows with the atomic
number and with c.m.\ energy; the radius $r_s$ of the cross section of a
string has been estimated in various ways 
\cite{Pajares:perc,NardiSatz} to be in the range
$0.2 - 0.3$~fm; in this paper we use $r_s = 0.2$~fm, 
and discuss the changes expected for larger values.

When two or more strings overlap even partially a larger structure is
formed in which colour can flow (following other authors
\cite{Pajares:perc}, we call such structures ``clusters of fused
strings''); clusters are shown shaded in Figure~\ref{fig:discs}.
When the distribution of strings in the plane is uniform,
the system is a typical example of the well studied problem of continuum
percolation in two dimensions \cite{Isichenko}: discs of a fixed
radius $r_s$
are randomly scattered in a plane following a uniform distribution.
Overlapping discs form clusters of different sizes, but
when the density of discs is large enough, an ``infinite cluster''
(a cluster that stretches to infinity in all directions) 
is formed with probability 1 and percolation is said to set in.
The minimum density at which the infinite
cluster is formed with probability 1 is known as ``percolation
threshold'': it corresponds to a phase transition whose critical
exponents are well studied. It is convenient to introduce a
dimensionless density $\eta$:
\begin{equation}
  \eta = n \pi r_s^2
\end{equation}
where $n$ is the number of discs per unit area. The percolation
threshold $\eta_c$ has been computed to be in the range 1.12--1.17
\cite{Isichenko,Pike74}. 
Thus the probability of having an infinite cluster is simply
\begin{equation}
  \Pp(\eta) = \theta(\eta-\eta_c)	\label{eq:theta}
\end{equation}

Coming back to the physical problem, it must be noticed that
there is no general agreement on the relationship 
between the density threshold for
percolation and the threshold for QGP formation:
it is widely believed that the latter
cannot be smaller than the former, and seems accepted by
most authors \cite{Pajares:perc,NardiSatz} that they are very close
to each other.
There is however one important point in which nucleus-nucleus
collisions differ from the uniform percolation model: the
distribution of strings in the transverse area is not uniform.

Indeed it is known that the
nuclear matter is more dense near the centre of the nucleus than
at the border, and in fact the density is usually parametrised 
in three dimensions with a Wood-Saxon distribution.
It is then clear that the probability of two partons interacting is
larger in the centre than at the border, and so is expected the
density of strings: to be higher in the centre, decreasing towards the border.
In the following, we explore how the percolation threshold
changes when the distribution of discs is not uniform.
The actual distribution could be approximated in several ways, 
making the problem very computation-intensive and model-dependent.
We prefer instead to address the issue by
comparing two limiting cases: the well known one of a uniform
distribution, and that of a Gaussian distribution.
The actual distribution is expected to lie between the two.

A computer simulation is performed in which a given number $N$ of discs 
(we treat in the following the terms `disc' and `strings' as synonyms)
of radius $r_s = 0.2$~fm is scattered over a circular surface of radius 
$R(A)$ where $A$ is the atomic number of the colliding nuclei, as
given e.g.\ by the usual formula:
\begin{equation}
  R(A) = 1.14\, A^{1/3}
\end{equation}
(when the two nuclei are not identical, we still take the impact
parameter to be zero, and the transverse area is then the transverse
area of the smaller nucleus).

In the standard case of continuum percolation, the discs are
uniformly distributed over the transverse area $0 \le r \le R(A)$;
the probability density in polar coordinates is given by:
\begin{equation}
  U(r,\phi) = \begin{cases}
	1/\pi R^2(A) & \text{if}~ r \le R(A)\\
	0&             \text{otherwise}
	\end{cases}					\label{eq:uniform}
\end{equation}
Of course, the dimensionless string density is given by:
$\eta = (r_s/R(A))^2 N$, $N$ being the number of strings.

\begin{figure}
  \begin{center}
  \mbox{\epsfig{file=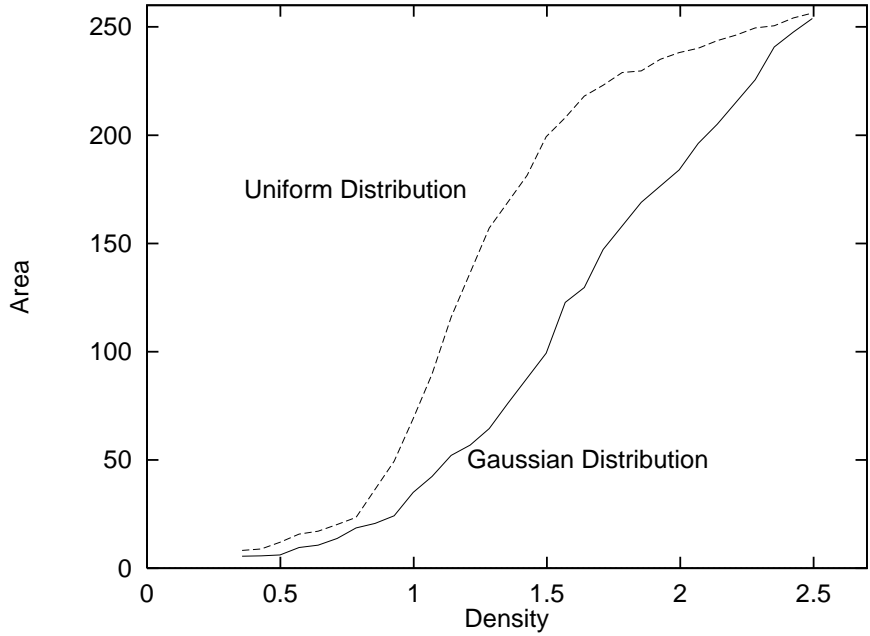,width=0.76\textwidth}}
  \end{center}
  \caption[area of largest cluster]{Geometrical area of the largest cluster
  $\Smax$ in $\text{fm}^2$
  (averaged over many simulations) versus dimensionless density $\eta$
  for the uniform distribution (dashed line) and for the Gaussian
  distribution (solid line).}\label{fig:Smax}
  \end{figure}

\begin{figure}
  \begin{center}
  \mbox{\epsfig{file=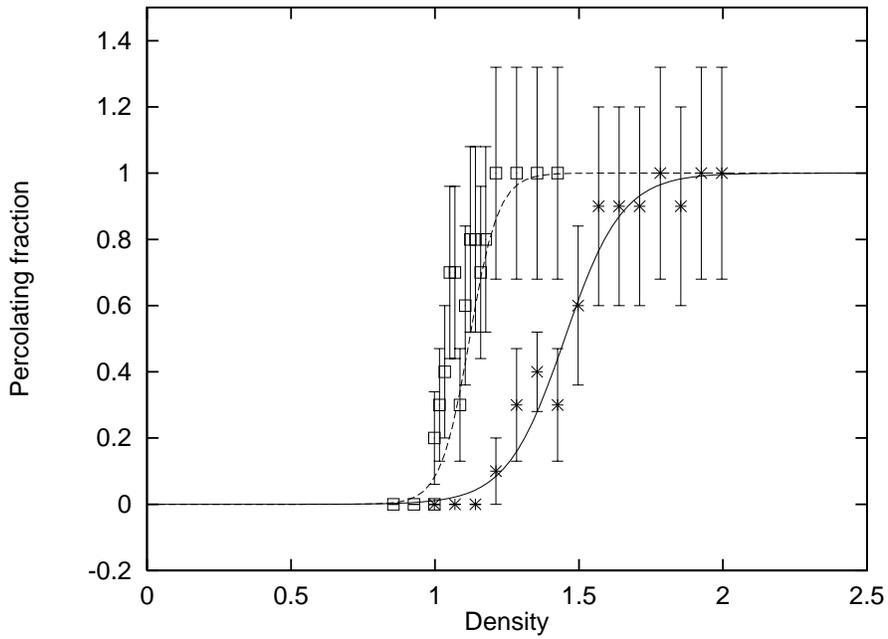,width=0.76\textwidth}}
  \end{center}
	%\vspace{5cm}
  \caption[probability of percolation]{Fraction of events in which
  percolation occurs, $\fp(\eta)$, vs. dimensionless density $\eta$.
	Open squares: uniform distribution; stars: Gaussian
  distribution. The lines show fits with the sigmoid 
  function. The error bars are purely statistical (Poissonian).
  }\label{fig:pprob}
  \end{figure}

Because we are dealing with a finite size system, we have to
approximate the usual percolation parameters with finite size
quantities.
The usual parameter is the probability
$P_\infty(\eta)$ for a disc to be part of the infinite cluster, which
behaves near the threshold as a power law:
\begin{equation}
  P_\infty(\eta) \approx (\eta - \eta_c)^\beta \theta(\eta-\eta_c)
		\qquad \text{for}~~ \eta \to \eta_c		\label{eq:threshold}
\end{equation}
where $\beta$ is the well known critical exponent \cite{Isichenko}.
Convenient parameters which approximate this behaviour are related to
the size of the largest cluster, for example $\Smax$,
the area occupied by it, \cite{Zeldovich}.
It is clear that this parameter is monotonically increasing,
but its increase below and above percolation threshold is much slower
than in the vicinity of $\eta_c$, where it grows very rapidly.
Figure~\ref{fig:Smax} (dashed line) shows exactly this behaviour
as obtained in our simulation.
Another simple definition of percolation, useful in the case of a
finite size system, is based on the diameter of a cluster, defined as
the largest of all distances between pairs of 
connected strings (see e.g. the
thick black lines drawn in the clusters of Figure~\ref{fig:discs})
\cite{Zeldovich,Pike74}.     When, in a particular
simulation, there exists a cluster whose diameter 
is comparable to the diameter of the available area, $2 R(A)$ 
in our case, we say that percolation ``occurs''. 
A good statistics is then the probability $\fp(\eta)$,
the fraction of events at given $\eta$ in which percolation occurs.
Its behaviour will approximate that of the probability of percolation
$\Pp(\eta)$, turning the Heavyside function of eq.~\eqref{eq:theta}
into a sigmoid function because of the finite size of the system.
Figure~\ref{fig:pprob} (open squares) shows the
results of our simulation for the uniform distribution. We can fit our
points very well with $\eta_c = 1.12$.

Let us now turn to the second interesting case:
the strings are distributed according to a two-dimensional Gaussian
distribution in the plane $r < \infty$:
\begin{equation}
  G(r,\phi) = \frac{1}{2\pi\sigma^2} \exp\left[ -
  \frac{r^2}{2\sigma^2} \right]		\label{eq:gauss}
\end{equation}
where we choose $\sigma = R(A)/\sqrt{2}$, so that the root-mean-square
radius equals the nuclear radius $R(A)$ and we can 
compare the results with those of the uniform distribution,
eq.~\eqref{eq:uniform}.
Because the density is higher in the centre and lower away from it,
we expect the percolation threshold in this case to be larger than for
the uniform distribution case.
Indeed in Figure~\ref{fig:Smax} we also show $\Smax$ for the Gaussian
case (solid line): while a precise determination of a threshold
for percolation in the Gaussian case is beyond the purpose of this
paper, it is nonetheless quite clear that this threshold lies above
the one for the uniform case.
Notice that this result is clearly confirmed by examining the
variation of the fraction of percolating events $\fp(\eta)$
in Figure~\ref{fig:pprob}:
from fitting this simulation we estimate $\eta_c \approx 1.5$ for the
Gaussian case.

\begin{figure}
  \begin{center}
  \mbox{\epsfig{file=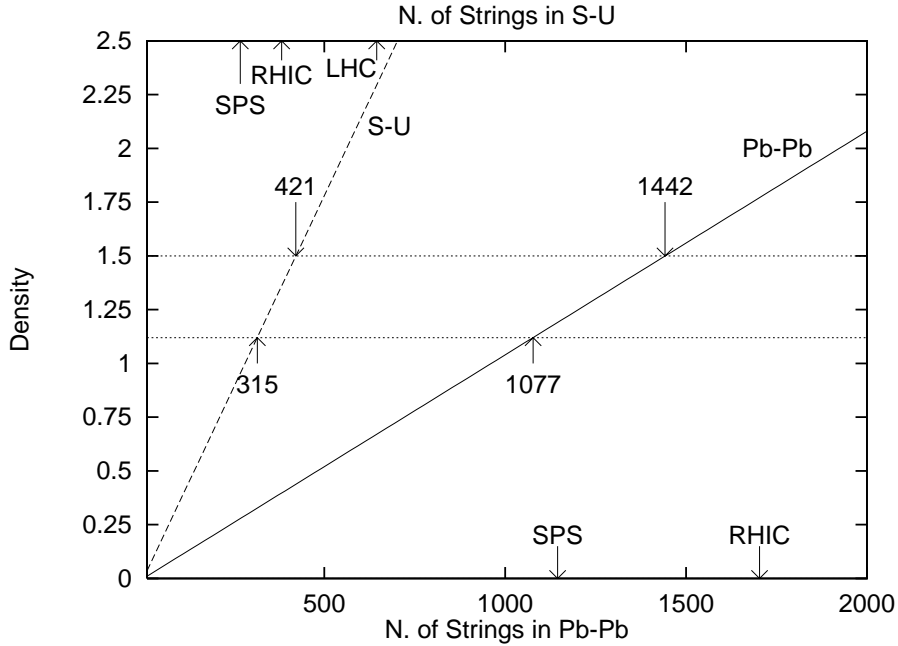,width=0.76\textwidth}}
  \end{center}
\caption[density vs number of strings]{Dimensionless density $\eta$
vs. number of strings in a collision for Pb-Pb
(solid line) and for S-U collisions (dashed line). The horizontal
lines mark the percolation thresholds for the uniform (lower line) and
for the Gaussian (upper line) distributions. Indicated is also the
relationship between the c.m.\ energy of current and future
experiments and the number of strings (after \cite{Pajares:perc}),
on the bottom axis for Pb-Pb and on the top axis for S-U.
}\label{fig:density}
\end{figure}

We apply now our results to nucleus-nucleus collisions.
In Figure~\ref{fig:density} we show the dimensionless density 
for Pb-Pb and S-U collisions as a function of the number of
strings created in the event.  Also shown are the percolation
thresholds for the uniform and Gaussian case. 
Even when the percolation threshold is identified with the
QGP formation threshold, one cannot conclude that present day
experiments have reached QGP.
This is in agreement with the results of a phenomenological
approach to rare events (like $J/\psi$ production) based on
scaling laws from the minimum bias distribution
\cite{DiasDeus:Jpsi}.
It should on the other hand be expected that the threshold for
percolation is reached at RHIC for Pb-Pb collisions, but it will
probably need the LHC for the lighter S-U collisions.

It must be said that Figure~\ref{fig:density} has been drawn with the
parameters of \cite{Pajares:perc} for the radii ($r_s = 0.2$ fm,
$R(\text{Pb}) = 6.2$ fm, $R(\text{S}) = 3.35$ fm). However the slopes
in the Figure are very sensitive to the choice of string radius.
A situation in which at present energy
the Pb-Pb collisions are above and S-U below
percolation threshold can arise for $r_s = 0.23$ fm. A larger choice,
like the one proposed in \cite{NardiSatz}, $r_s = 0.26$, fm 
will put S-U collisions slightly above percolation threshold already
at the SPS collider.

\section*{Acknowledgements}
This work was partially supported by project PRAXIS/PCEX/P/FIS/124/96.
R.~Ugoccioni would like to acknowledge the financial support of the
``Sub-Programa Ciência e Tecnologia do $2^o$ Quadro 
Comunitário de Apoio.''

\section*{References}

%%%% References file: percolation.ref
%%%% Creator: REFLATEX.PL 1.0
%%%% Usage: \input{percolation.ref} where references should go.
%%%%

\end{document}